\documentstyle[aps,preprint,floats,psfig]{revtex} 

\def\tev{\rm TeV} 
 
\def\lsim{\mathrel{\raise.3ex\hbox{$<$\kern-.75em\lower1ex\hbox{$\sim$}}}} 
\def\gsim{\mathrel{\raise.3ex\hbox{$>$\kern-.75em\lower1ex\hbox{$\sim$}}}} 

\newcommand{\mpl}{M_{\text{Pl}}}
\newcommand{\ms}{M_S^{}}
\newcommand{\mssq}{M_S^{2}}
\newcommand{\nul}{\nu_L^{}}

\begin{document} 
 

\tighten
\preprint{
\vbox{\hbox{\bf MADPH-02-1262}
      \hbox{\bf hep-ph/0204112}
      \hbox{September, 2002}}}

\draft 
\title {TeV String State Excitation via High Energy Cosmic Neutrinos} 
\author{Joshua J. Friess, Tao Han, Dan Hooper} 
\address{   
Department of Physics, University of Wisconsin, 1150 University Avenue,   
Madison, WI 53706 
} 
\date{April, 2002} 
 
\maketitle 

\begin{abstract}
We construct the open-string scattering amplitudes
for neutrino-gluon collisions and evaluate the high energy
neutrino-nucleon scattering cross section via string state excitations
in the TeV string-scale scenario. 
We find that the neutrino-gluon scattering is the dominant
contribution, $5-10$ times larger than neutrino-quark processes, 
though black hole production may be larger 
than the string contribution at higher energies.  
We illustrate the observability of the string signal at the Auger 
Observatory and the IceCube neutrino telescope
for the string scale $\ms\simeq 1$ TeV.
\end{abstract}

\section{Introduction}

Theories with a low fundamental string scale ($\ms$) \cite{lowscale}
may help explain the apparent mass hierarchy between the electroweak
scale and the Planck scale ($\mpl$) \cite{add,rs}. These
interesting new scenarios may also lead to rich low energy 
phenomenology \cite{pheno} that may be observable
in near-future high energy experiments \cite{addpheno}.
Generally speaking, at an energy scale probing a distance smaller
than the size of extra dimensions, one expects to
explore the nature of higher dimensional physics, where gravitons 
propagating in the bulk play 
an important role in the dynamics \cite{add}. 
At energies much larger than the fundamental scale $M_S$, 
black hole production with semi-classical treatments
may be the dominant phenomenon \cite{bh,bhs}.
At energies near the string scale threshold, however,
the physics may be more involved, and the string resonance
states should become very important \cite{stringy}.
Perturbative arguments show that the (open) string 
scattering amplitudes should be dominant over 
the (closed string) graviton processes \cite{stringy}.

In this paper, we study the TeV string state excitations 
via high energy cosmic neutrinos. The motivation is multi-fold.
First, high energy cosmic rays provide a natural accelerator.
Neutrinos are not subject to the GZK cutoff and
may arrive at the Earth with ultra-high energies, leading to
potentially spectacular events \cite{neus} 
in the atmosphere or in the Earth's crust. Second, models
with a low string scale typically predict enhanced neutrino
interactions with matter at higher energies \cite{shrock,doug}. 
For instance, an effective field theory calculation for large 
extra dimension scenarios leads to a $\nu g$ scattering amplitude 
${\cal M}\sim s^2/M_S^4$ \cite{doug}, where $s$ is the 
c.m.~energy squared. This energy growth, 
however, violates partial wave unitarity near the
string scale, thus invalidating the perturbative treatment.
On the other hand, the open-string scattering amplitudes, 
which are manifestly unitary, should be the appropriate 
description for physics in this energy regime \cite{stringy,cpp}. 
In fact, the neutrino-nucleon scattering
cross section has been constructed in this scenario \cite{cim},
and the physical consequences have been studied in the context 
of the km-squared neutrino telescope \cite{fred}.
However, the scattering amplitudes were constructed only for
neutrino-quark interactions. Since we are exploring the
new physics typically at energies of a TeV, one expects 
that gluons in a nucleon would become the dominant partons
to participate the interaction.
Finally, it is important to explore the relative contributions 
from string excitations and black hole production.
In the next section, we construct the open-string  
amplitudes for neutrino-gluon $\nu g$  scattering.
We then study the string resonances in the $s$-channel
and evaluate the neutrino-nucleon $(\nu N)$ scattering 
cross section.
We compare the string state contribution with that of
black hole production, and finally, we demonstrate the signal 
observability at the Pierre Auger
air-shower array observatory, as well as the IceCube neutrino telescope.

\section{ $\nul g$ Elastic Scattering Amplitude}

The general tree-level open-string amplitudes can be expressed
by \cite{ampl}
\begin{eqnarray}
 {\cal M}(1,2,3,4) =  g^2 [A_{1234} \cdot S(s,t) \cdot T_{1234}
  +    A_{1324} \cdot S(t,u) \cdot T_{1324}  
  +    A_{1243} \cdot S(s,u) \cdot T_{1243}\ ]\ ,
\label{FA}
\end{eqnarray}
where $(1,2,3,4)$ denote the external states with momenta directed 
inward. $A_{ijkl}$ are the kinematic factors for the color-ordered
amplitudes \cite{mp}, associated with the expansion basis $T_{ijkl}$, 
the so-called Chan-Paton factors.
$S(s,t)$ is essentially the Veneziano amplitude \cite{ampl}
\begin{equation}
S(s,t) = \frac{\Gamma(1 - \alpha's)\Gamma(1 - \alpha't)}
{\Gamma(1 - \alpha's - \alpha't)}\ ,
\label{vene}
\end{equation}
where $\alpha'=1/{\ms}^2$ is the Regge slope. 
In the zero-slope limit $\alpha's\to 0$, the Veneziano
amplitudes approach unity, and Eq.~(\ref{FA})
should reduce to the field theory result at energies 
far below the string scale.

Identifying $(\nu_L, g, g, \nu_L)\to (1,2,3,4)$, 
we can easily derive the relevant factors for our amplitudes.
For instance, for a right-handed gluon, we have \cite{cpp,mp}
\begin{eqnarray}
 A_{1234} = -4\frac{u}{t} \sqrt{\frac{u}{s}},\ 
 A_{1324} = -4\frac{s}{t} \sqrt{\frac{u}{s}},\ 
 A_{1243} = -4\sqrt{\frac{u}{s}}\ .
\nonumber 
\end{eqnarray}
Substituting these back into Eq.~(\ref{FA}) and taking
the (3,4) momenta outgoing, we get the full amplitude:
\begin{eqnarray}
\label{intm_right} 
{\cal M}(\nu_L g_R \rightarrow \nu_L g_R) = 
-4 g^2 \frac{1}{t} \sqrt{\frac{-u}{s}}\ 
 [u S(s,t) T_{1234} + s  S(t,u) T_{1324} + t S(s,u) T_{1243}]\ .
\nonumber
\end{eqnarray}
For a model given by embedding 
the SM fields into a D-brane structure in string theories, 
one will be able to calculate explicitly the Chan-Paton $T$ factors. 
For a typical embedding into a $U(N)$ gauge group
(with the group generators normalized 
to $Tr(t^a t^{b\dagger})={1\over 2}\delta^{ab}$),
the Chan-Paton factors can be $T \sim {1/4}$ to 1.
For instance, in a QED toy model \cite{cpp}, the Chan-Paton factors
for $e^-_Le^+_R\to \gamma_L\gamma_R$ are $T=\pm 1/4$.
As a consistency check of this method, Eq.~(\ref{intm_right})
leads to the familiar QED Compton scattering amplitude
when the Chan-Paton factors are evaluated
to be $T_{1234} = T_{1324} = -T_{1243} =1/4$,
following \cite{cpp}.

On the other hand, explicit constructions of string embedding
often lead to new states that are incompatible with the SM
at low energies. Instead, we take a rather 
model-independent approach:
Although we do not assign an explicit representation of
the $U(N)$ generators to the SM particles, 
we can parameterize the $T$'s by requiring that Eq.~(\ref{intm_right}) 
reproduce the SM result at low energies.  
Since the tree-level SM result for the $\nu g$ scattering 
vanishes and the Veneziano amplitudes approach unity
at low energies ($\alpha's\to 0$), we require
\begin{equation}
0 = u\ T_{1234} + s\ T_{1324} + t\ T_{1243}\ .
\end{equation}
This relation is satisfied in the massless limit if and only 
if $T_{1234} = T_{1324} = T_{1243} \equiv T$. 
The corresponding amplitude for a left-handed gluon
can be obtained in a similar way.
We thus arrive at our final expressions:
\begin{eqnarray}
&& {\cal M}(\nu_L g_R \rightarrow \nu_L g_R) = 
-4 g^2 T\ \frac{1}{t}\ \sqrt{\frac{-u}{s}}\times
\left[u S(s,t) + s S(t,u) + t S(s,u)\right], 
\label{final_right} \\
&& {\cal M}(\nu_L g_L \rightarrow \nu_L g_L) = 
{\cal M}(\nu_L g_R \rightarrow \nu_L g_R)_{u\leftrightarrow s}\ .
\label{final_left}
\end{eqnarray}
Since we are uninterested in the particular
color structure in practice, and for simplicity of 
the presentation, we have taken the Chan-Paton factors
for different gluons to be equal.
The scattering amplitude is completely specified by the string
coupling $g$, string scale $\ms$, and a model parameter $T$.

\section{ String Resonances and the $\nu N$ Cross Section}

It is informative to understand the physical interpretation
of the $\nu g$ amplitude constructed above. 
The Veneziano amplitude of Eq.~(\ref{vene})
presents $s$-channel poles, as well as $t$- and $u$-channel
exchanges. Obviously, the $s$- and $u$-channels are due to
exchanges of fermionic states of (${\bf 8,2}$) under
$SU_C(3)\otimes SU_L(2)$, called lepto-gluons ($\nu_8$). 
The $t$-channel involves new exotic bosons which may 
be in the adjoint representations of the SM gauge groups.

Although our amplitudes
of Eqs.~(\ref{final_right}$-$\ref{final_left}) reduce 
to the SM results at $s\ll \mssq$, at energies near or 
above the string threshold, 
the string resonances in the $s$-channel dominate. 
The Veneziano amplitude has simple poles at $s = n\mssq$
for any positive integer $n$,
\begin{eqnarray}
\nonumber
S(s,t) \approx \mssq\ \sum_{n=1}^{\infty} \frac{(t/\mssq)(t/\mssq +1)
\cdots(t/\mssq + n - 1)}{(n-1)!(s-n\mssq)}\ .
\end{eqnarray}
Off-resonance contributions to the cross section
are negligible. There are also terms proportional to $S(t,u)$.
We will neglect these terms and approximate the amplitude 
near its $s$-channel poles \cite{cim}. 
The amplitude for right-handed gluons now becomes
\begin{eqnarray}
{\cal M}(\nu_L g_R \rightarrow \nu_8 \rightarrow \nu_L g_R) \approx 
\sum_{n=1}^{\infty} A_n\ ,
\end{eqnarray}
where
\begin{eqnarray}
\nonumber
A_n &=&
\frac{-4 g^2 T \mssq }{(n-1)!(s-nM^2_S)}\ \frac{1}{t}\
\sqrt{\frac{-u}{s}} \times\ \left[u\frac{t}{M^2_S} (\frac{t}{M^2_S} + 1)
\cdots(\frac{t}{M^2_S} + n -1) + 
( t\leftrightarrow u)\right]\\
\nonumber
&=&
\frac{-8 g^2 T \mssq }{(n-1)!(s-nM^2_S)}\ \frac{u}{t}\ 
\sqrt{\frac{-u}{s}}\times\  \frac{t}{M^2_S}(\frac{t}{M^2_S} + 1)
\cdots(\frac{t}{M^2_S} + n -1)\quad  \mbox{for odd $n$}
\label{wacky_sum}
\end{eqnarray}
and $A_n=0$ for even $n$. It is convenient to 
expand the amplitude \cite{cim} in terms of the 
Wigner functions $d^J_{mm'}$. Since both initial and final states 
for $\nu_L g_R \rightarrow \nu_L g_R$ have total helicity
$J_Z = 3/2$, the result reads
\begin{equation}
A_n = \frac{8g^2 T nM^2_S}{s-nM^2_S} \mbox{ } \sum_{J = 3/2}^{n+1/2} 
\alpha_n^J d_{3/2,3/2}^J\ ,
\end{equation}
where the coefficient $\alpha_n^J$ satisfies 
the normalization relation
$\sum_{J = 3/2}^{n+1/2} \left|\alpha_n^J\right| = 1$.

Equipped with the amplitude for 
$\nu_L g_R \to (\nu_8)_n^J \rightarrow \nu_L g_R$, 
we can determine the partial decay width for 
$(\nu_8)_n^J \rightarrow \nu_L g_R$:
\begin{equation}
\Gamma_n^J \equiv \Gamma((\nu_8)_n^J \rightarrow \nu_L g_R) = 
\frac{g^2}{2\pi}\frac{|T|}{2J+1}\sqrt{n}M_S \left|\alpha_n^J\right|\ .
\end{equation}
In the narrow-width approximation, this leads to
the cross section $\sigma(\nu_L g_R \rightarrow (\nu_8)_n^J)$
\begin{eqnarray}
\nonumber
\sigma_n^J(\nu_L g_R) & = & \frac{4\pi^2 \Gamma_n^J}
{\sqrt{n}M_S} (2J+1)\delta(s - nM_S^2)
 = 2\pi g^2 \left|T \alpha_n^J\right| \delta(s-nM_S^2)\ .
\end{eqnarray}
We now sum these partial cross 
sections for a fixed odd $n$, obtaining
\begin{equation}
\sigma_n \equiv \sum_J \sigma_n^J(\nu_L g_R) = 
\tilde{\sigma}_n \delta(1 - nM^2_S/s)\ ,
\label{sigman}
\end{equation}
where $\tilde{\sigma}_n \equiv 2\pi g^2 |T|/nM^2_S$.
The partial cross sections for left-handed gluons turn out to be 
identical to Eq.~(\ref{sigman}).

Finally, the neutrino-nucleon deeply inelastic
scattering cross section is given by
\begin{equation}
\sigma(\nu_L N) = \sum_{n=1}^{n_{cut}} \sum_{f} 
\tilde{\sigma}_n (\nu_L f)\ x f(x,Q^2)\ ,
\end{equation}
where $x = nM_S^2/S$ is the energy fraction carried by a parton;
$S$ is the neutrino-nucleon c.m.~energy squared, and
$Q^2$ is taken to be $nM_S^2$. The sum $f$ is over all 
contributing partons in the nucleon.
Quark parton contributions were evaluated in \cite{cim}.
As part of our current motivation, we expect that the
gluonic contribution as constructed above will dominate at high energies.

\begin{figure}[thb] 
\centerline{\psfig{file=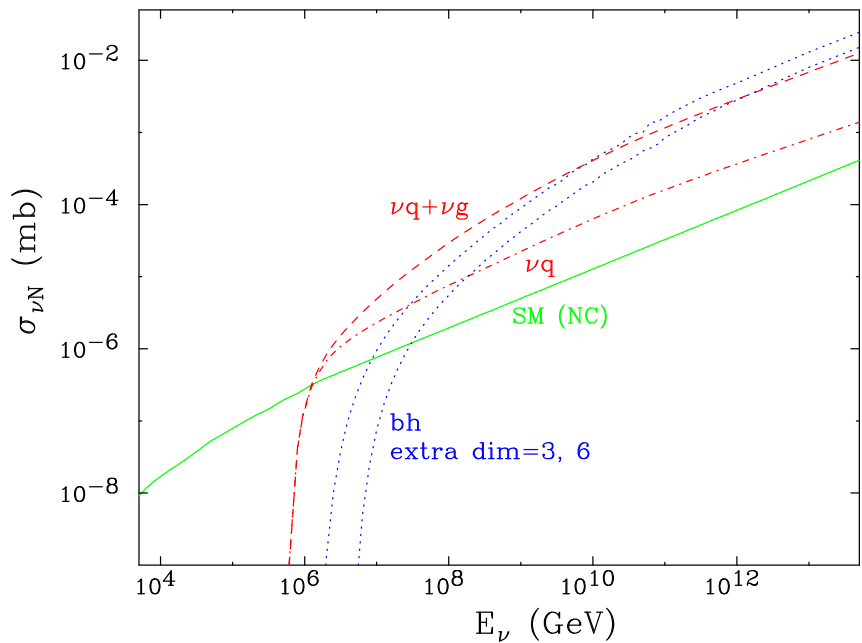,width=4.5in}}
%
\caption{$\nu N$ cross sections via TeV string resonances
from $\nu q+\nu g$ contribution (the dashed curve)
and $\nu q$ only (the dot-dashed).
The string scale is taken to be $\ms$=1 TeV and the 
Chan-Paton factor is $T=1/2$.
Also plotted are the SM neutral current prediction (solid curve)
and the black hole production (dotted) with 3 (upper) and 6
extra dimensions. }
\label{figure1} 
\end{figure}

We present our results in Fig.~\ref{figure1}. The solid curve
shows the prediction for the SM neutral current process.
We have used the CTEQ5-DIS parton distributions, 
extended to $x < 10^{-5}$ using the methods in \cite{hallsie}.
The neutrino-nucleon cross sections via 
string excitations with and without the gluon contribution 
are depicted by the dashed and dot-dashed curves, respectively.
Throughout our presentation, we have taken the string coupling $g=1$, 
which keeps the effective coupling $g^2/4\pi$ perturbative.
For comparison, we have taken $T = 1/2$, 
which puts the Chan-Paton trace factor for the gluons
to be the same as that for the quarks when $a = b = 5$ 
in the parameterization of \cite{cim}.
We have only summed over the resonances to $n_{cut}=50$.
Our results are rather insensitive to the choice of $n_{cut}$. 
For example, taking $n_{cut} = 20\ (80)$ instead 
reduces (increases) our final results by about $10\%$.
The new physics threshold starts near $E_\nu\sim 10^3$ TeV,
which corresponds to our choice of $\ms=1$ TeV. 
Due to the large gluon luminosity at high energies,
including the gluons in the total cross section increases 
the final result by about a factor of $5-10$ for 
$E_\nu\sim 10^5-10^{10}$ TeV.

As commented in the introduction, graviton exchange in the
large extra dimension scenario is the leading effect below
$\ms$ \cite{pheno,fred}.  However, black hole production far
above $\ms$ may be dominant \cite{bh,bhs}. We show the 
black hole production
with the geometric cross section \cite{bhs}
for $s\ge \ms=1$ TeV by the dotted curves with
3 (upper) and 6 large extra dimensions.\footnote{When comparing with
black hole production, we have 
adopted a convention to relate our string scale and $(4+n)$-dimensional 
gravity scale as $\ms=[8\pi/(2\pi)^{n}]^{1/n+2}M_D^{}$. Numerically,
It leads to $\ms=0.63M_D\ (0.38M_D)$ for $n=3\ (6)$. In other words,
a choice of $\ms=1$ TeV corresponds to $M_D=$ 1.6 (2.6) TeV for $n=3$ (6).}
Indeed, the black hole cross section may take over for  
$E_\nu\gsim 10^6$ TeV.
We finally note that
the cross section of the order $\mu$b at $E_\nu\sim10^{20}$ eV 
is much too low to account for the ultra-high energy 
cosmic ray events that violate the GZK bound \cite{neus}.

\section{ Signatures in Air-shower Cosmic Ray Experiments and 
in Neutrino Telescopes}

At energies inaccessible to colliders, astroparticle physics 
experiments \cite{astro1,astro2} may provide a window into new physics. 
Enhanced neutrino-nucleon cross sections may be an observable 
signature for theories with a low string 
scale \cite{shrock,doug,cim,fred,others}.

Very high energy neutrinos are predicted to be generated in a variety of
astrophysical sources. We consider two of these possibilities: neutrinos
from compact sources and cosmogenic neutrinos.  For the flux from compact
sources, we use the limit placed by Waxman and Bahcall of 
$E_\nu^2 dN_\nu/dE_\nu=2\times 10^{-8}\ \rm{GeV}\ 
\rm{cm}^{-2}\rm{s}^{-1}\rm{sr}^{-1}$ \cite{wb}.  This flux describes
neutrinos from sources such as gamma-ray bursts and blazars which are
assumed to also generate the highest energy cosmic rays.  The cosmogenic
neutrino flux is generated by cosmic rays scattering off of the cosmic
microwave background \cite{stecker}.  We use the flux as calculated by
Seckel {\it et al.} \cite{seckel}.  This flux peaks at
$E_\nu\sim 10^9$ GeV and is smaller than the
Waxman-Bahcall limit over all, but is also a more robust prediction.  
We stress that determining the neutrino flux is one of the primary 
goals for the neutrino telescope experiments \cite{astro1,astro2}
and should be the first step toward searching for new physics.
We consider two classes of next generation high energy astroparticle
physics experiments: the Auger air-shower array \cite{auger} and 
the kilometer-scale neutrino 
telescope IceCube \cite{nature}.

At EeV energies, neutrinos can generate 
atmospheric air showers observable in cosmic ray experiments.  
The event rates at these extremely high energies 
scale as the cross section of 
interacting neutrinos and thus can be sensitive to new physics 
that enhances the cross section \cite{astro1}. 
We present our numerical results in Table \ref{table:I}
for the quasi-horizontal events with a zenith angle larger than 
$75^\circ$, including the prediction from the SM interactions
and those from the string resonances. 
It is important to note that the two neutrino
fluxes considered only lead to a difference of a factor of $2-3$ 
at these very high energies,
making the uncertainties due to the unknown flux less severe. 
With a few years of data taking, it is conceivable to establish 
a signal for $\ms\sim 1$ TeV statistically over the SM 
expectation at the Auger Observatory.

Neutrino telescopes, optimized for TeV-PeV neutrinos, 
can measure neutrino-nucleon cross sections given a 
sufficiently high energy neutrino flux.  This is done by comparing 
the energy and angular distributions of showers generated in the 
detector to SM predictions. As cross sections increase at higher 
energies, the Earth becomes opaque to neutrinos, thus suppressing the
up-going neutrino event rate.  Down-going rates, however, become 
further enhanced.
The ratio of up-going to down-going events is an effective
measurement for the cross section at a given energy \cite{astro2}. 
At IceCube, we consider a shower threshold energy of
$E_{\rm{sh}}^{\rm{th}}=250$ TeV (corresponding to a neutrino 
energy near 1 PeV) to enhance the signal to background. 
We see again that with a
few years of data taking, a signal above the SM expectation
may be identifiable at IceCube. Unlike for very high energy 
air-shower experiments, the two fluxes considered lead to substantially
different event rates, due to the fact that the Waxman-Bahcall
flux is much larger at these lower energies. We can thus see the
complementarity of the Auger Observatory and IceCube.

\begin{table}[thb]
\begin{tabular}{ c|c c c c c c } 
Auger (events/yr) & 
\multicolumn{2}{c }{~WB Flux~} & \multicolumn{2}{c }{~Cosmogenic
Flux~} \\ 
\hline \hline
~~SM ($E_{\rm{sh}}^{\rm{th}}=10$ PeV)
~~&~0.66~&~~&~0.20~&~~\\
\hline
~~$M_S=1~\tev$ 
~~&~3.15~&~~&~1.25~&~~\\
~~$M_S=2~\tev$ 
~~&~0.96~&~~&~0.34~&~~\\  \hline  \hline \hline 
& \multicolumn{2}{c }{~WB Flux~} & \multicolumn{2}{c }{~Cosmogenic
Flux~} \\ \cline{2-5}
\raisebox{1.7ex}[0pt]{~IceCube (events/yr)} 
&~Down~~&~Up~~&~Down~~&~Up~~\\ 
\hline \hline
~~SM ($E_{\rm{sh}}^{\rm{th}}=250$ TeV)
~~&~8.4~&~1.8~&~0.15~&~0.012~\\
\hline
~~$M_S=1~\tev$ 
~~&~14.8~&~2.2~&~0.72~&~0.024~\\
~~$M_S=2~\tev$ 
~~&~8.7~&~1.9~&~0.22~&~0.016~
\end{tabular}
\caption{The SM and string excitation event rates 
for ultra-high energy quasi-horizontal air showers 
in the Auger Observatory, and 
down-going and up-going showers (in $2\pi$ sr) in IceCube.  
The Waxman-Bahcall~\protect\cite{wb} and
cosmogenic~\protect\cite{seckel} fluxes are considered.
The Chan-Paton factor is taken as $1/2$.}
\label{table:I} 
\end{table}

\section{ Summary}

We constructed the neutrino-gluon amplitudes as open-string 
scattering, and subsequently studied neutrino-nucleon collisions 
via string state excitations in the low string scale scenario.
We found that neutrino-gluon scattering is the dominant
process, $5-10$ times larger in rate than the neutrino-quark
induced processes \cite{cim}. 
It is interesting to note that even for processes that vanish in 
the SM at tree-level, there can still be
substantial stringy contributions to their amplitudes 
at high energies. We also demonstrated that black hole 
production may be larger than the string state
contribution at higher energies.  
Finally, we evaluated the signal rates at the Auger Observatory 
and the IceCube neutrino telescope and illustrated
the possibility of observing the signal if the string scale 
is near 1 TeV.

\vskip 0.2cm

{\it Acknowledgments}: We thank Jonathan Feng, Maxim Perelstein, 
and Hallsie Reno for helpful discussions. 
This work was supported in part by a DOE grant No.  
DE-FG02-95ER40896 and in part by the Wisconsin Alumni Research Foundation.

\end{document}